\documentclass[journal=jacsat,manuscript=article]{achemso}

\usepackage[version=3]{mhchem} 

\usepackage[]{hyperref}
\usepackage{graphicx}
\usepackage{amsmath}
\usepackage[normalem]{ulem}
\usepackage{siunitx}
\usepackage{soul}
\usepackage{amssymb}
\makeatletter
\usepackage{rotating}
\usepackage{url}
\usepackage{anyfontsize}
\usepackage{blindtext}
\usepackage[utf8]{inputenc}
\usepackage{xr}
\usepackage{lipsum}

\usepackage[font= large,label font=bf,labelformat=simple]{subfig}

\usepackage[T1]{fontenc}
\usepackage{lipsum}
\usepackage[table]{xcolor} 
\usepackage{tabularx}

\author{Sumit Sinha}
\affiliation{Department of Physics, University of Texas at Austin, Austin, TX 78712, USA.}
\author{Abdul N Malmi-Kakkada}
\email{amalmikakkada@augusta.edu*}
\affiliation{Department of Chemistry and Physics, Augusta University, Augusta, GA 30912, USA.}

\title[An \textsf{achemso} demo]
  {Inter-particle adhesion regulates the surface roughness of growing dense three-dimensional active particle aggregates}

\abbreviations{IR,NMR,UV}
\keywords{American Chemical Society, \LaTeX}

\begin{document}







\begin{abstract}
 Activity and self-generated motion are fundamental features observed in many living and non-living systems. Given that inter-particle adhesive forces are known to regulate particle dynamics, we investigate how adhesion strength controls the boundary growth and roughness in an active particle aggregate. Using particle based simulations incorporating both activity (birth, death and growth) and systematic physical interactions (elasticity and adhesion), we establish that inter-particle adhesion strength ($f^{ad}$) controls the surface roughness of a densely packed three-dimensional(3D) active particle aggregate expanding into a highly viscous medium.  
We discover that the surface roughness of a 3D active particle aggregate increases in proportion to the inter-particle adhesion strength, $f^{ad}$. We show that  asymmetry in the radial and tangential active particle mean squared displacement (MSD) suppresses 3D surface roughness at lower adhesion strengths. 
By analyzing the statistical properties of particle displacements at the aggregate periphery, we determine that the 3D surface roughness is driven by the movement of active particle towards the core at high inter-particle adhesion strengths. 
Our results elucidate the physics controlling the expansion of adhesive 3D active particle collectives into a highly viscous medium, with implications into understanding stochastic interface growth in active matter systems characterized by self generated particle flux. 
\end{abstract}

\section{Introduction}
Active matter systems constitute a type of matter whereby each unit particle utilizes energy 
to generate work~\cite{ramaswamy2010mechanics,marchetti2013hydrodynamics,gompper2020active}. There are many examples of active matter systems from both living~\cite{groswasser2018living} - cells, animals etc - and non-living contexts~\cite{vicsek2012collective} (molecules, droplets, grains etc) in the world around us. 
The energy required for movement can either be self-generated (as in the case of cells in biological tissues which convert stored energy into motion) or as in driven granular systems supplied externally by a shaking plate. As each particle (or `agent') utilizes energy at the individual level, active matter systems are out of equilibrium at the single particle scale~\cite{liesbeth2019topical}.  

The field of active matter originated from efforts to understand the physical underpinnings of collective behavior in animals~\cite{berthier2019glassy,vicsek1995}. Pioneering work by Vicsek and co-workers~\cite{vicsek1995} showed that the collective dynamics of self-driven particles emerge from a form of inter-particle coupling: a simple rule that the direction of motion of a particle is aligned  with the average direction of motion of its neighboring particles. As of yet it is unclear whether active particles universally possess such an intrinsic tendency to align its direction of motion with its neighbors~\cite{berthier2019glassy}. In living active matter, such as cells, adhesive inter-particle interactions can modulate the mechanical contact between particles thereby controlling its spatial structure and dynamics~\cite{maitre2012science,friedl2017tuning,pascalis2017single}. In the context of active colloids,  
selective and directional ligand mediated bonds can tune its motion and spatial configuration~\cite{wang2020ncomms}. 
Therefore, inter-particle adhesive interactions function as a crucial regulator of the collective dynamics and spatial 
structure of active matter systems. 

An important feature of living active matter is the presence of birth and death where the ratio of birth and death rates can lead to three distinct implications on the overall system size or total number of particles: (i) when  birth rate is less than the death rate, (ii)  birth and death rates are balanced and (iii) when birth rate is in excess of the death rate. In this work we focus on the third scenario where birth rate exceeds the death rate leading to a fast expanding collection of active particles with boundary growth. 
Study of growth and fluctuation of interfaces between one medium and another have laid the groundwork for important advances in physics. For instance, the Kardar-Parisi-Zhang(KPZ) model describes the stochastic growth profile of interfaces~\cite{kardar1986dynamic},  elucidating the behavior of materials ranging from crystal growth in thin films, turbulent liquid crystals to bacterial colony growth~\cite{allen2019bacterial}. In the KPZ model, the particle flux is orthogonal to the growing interface whereas another interesting scenario deals with particle flux  generated by the surface itself as is relevant in biological systems such as membranes and cell collectives~\cite{podewitz2016interface,williamson2018stability,patteson2018propagation}. In this context, Risler et. al.~\cite{risler2015homeostatic} described the out-of-equilibrium surface fluctuations of cell collectives in the homeostatic state when cell birth and death are balanced. Building on our prior work~\cite{malmi2018cell, malmi2019dual, sinha2020spatially, sinha2020self, samanta2020far} where we modeled biological cells with pairwise elastic and adhesive interaction in addition to rules for size growth, division and death of particles, we study  the surface roughness of a dense and fast expanding three dimensional (3D) collection of active particles. We focus on investigating 3D aggregate expansion at varying inter-particle adhesion strengths, known to critically tune collective properties in active matter systems. Our study models a dense 3D active particle aggregate expanding into a highly viscous surrounding medium, under free boundary conditions in the context of an off-lattice model. 

Our theoretical prediction and subsequent analysis of experimental data showed that active particles in the core versus the periphery of a dense 3D active particle aggregate exhibit dramatic variations in the dynamics~\cite{malmi2018cell, sinha2020spatially}. Whereas cells near the core are characterized by 
subdiffusive glassy dynamics (mean square displacement, $\Delta(t) \sim t^{\alpha}$ with $\alpha< 1$ ), cells in the
periphery exhibit highly directed superdiffusive motion, $\Delta(t) \sim t^{\alpha}$ with $\alpha > 1$. Such topological differences in the motion of constituent particles in a dense collection of active particles is reminiscent of the variation in the dynamics of particles between the surface and bulk in glasses~\cite{stevenson2008jcp}. Here, we address the question of how the emergent spatial heterogeneity in the dynamics of three-dimensional (3D) active particle aggregate determine the surface roughness during aggregate expansion into a highly viscous medium. We observe that the strength of inter-particle adhesive interactions strongly up-regulate the surface roughness of expanding dense 3D active particle aggregates. We show that the uptick in the 3D surface roughness is due to the heterogeneity in the radial displacements of the particles on the surface of the expanding 3D aggregates, driven by the symmetry between radial and transverse motions.

\section{Simulation details}
We briefly describe the simulation scheme adapted from our previous work on 3D tumor growth \cite{malmi2018cell, malmi2019dual, sinha2020spatially, sinha2020self, samanta2020far}. In the present study, we performed an off-lattice simulation of a growing 3D dense active particle aggregate~\cite{drasdo2005single, schaller2005multicellular}. In the growing aggregate (see Appendix, Movies 1-3), individual particles are modeled as soft deformable spherical agents. The individual particles grow stochastically in time and undergo division into daughter agents on reaching a critical size. The physics of the aggregate growth is governed by two factors - (a) systematic mechanical forces arising from two body interactions, (b) active processes due to particle growth, division and death, as we explain further below. 

{\bf (a) Systematic Interactions:} The individual particles interact with short-ranged forces, consisting of two terms, elastic force (repulsion) and adhesion (attraction). The elastic force ($F_{ij}^{el}$) between two particles $i$ and $j$ of radii $R_i$ and $R_j$ is given by
\begin{equation}
F_{ij}^{el}=\frac{h_{ij}^{3/2}}{\frac{3}{4}(\frac{1-\nu_i^2}{E_i}+\frac{1-\nu_j^2}{E_j})\sqrt{\frac{1}{R_i(t)}+\frac{1}{R_j(t)}}},
\end{equation}
where $\nu_i$ and $E_i$ are the Poisson ratio and elastic modulus of the $i^{th}$ particle. Also, $h_{ij}$ is the virtual overlap distance between the two particles. The adhesive force ($F_{ij}^{ad}$) is given by, 
\begin{equation}
F_{ij}^{ad}=A_{ij}f^{ad}\frac{1}{2}(c^{rec}_i c_j^{lig}+c^{lig}_i c_j^{rec}),
\end{equation}
where $A_{ij}$ is the overlap area between the two interacting particles and $f^{ad}$ determines the strength of adhesive bond. We have normalized the receptor(rec) and ligand(lig) concentration to satisfy $c^{rec}_i=c^{lig}_i=1$. 

\begin{table}
\begin{tabular}{ |p{6.5cm}||p{4cm}|p{1cm}|  }
 \hline
 \bf{Parameters} & \bf{Values}  \\
 \hline
 Timestep ($\Delta t$)& 10$\mathrm{s}$  \\
 \hline
Critical Radius for Division ($R_{m}$) &  5 $\mathrm{\mu m}$ \\
 \hline
Environment Viscosity ($\eta$) & 0.005 $\mathrm{kg/ (\mu m~s)}$  \\
 \hline
 Benchmark Cell Cycle Time ($\tau$)  & 54000 $\mathrm{s}$\\
 \hline
 Adhesive Coefficient ($f^{ad})$&  $0-3\times 10^{-4} \mathrm{\mu N/\mu m^{2}}$\\
 \hline
Mean Cell Elastic Modulus ($E_{i}) $ & $10^{-3} \mathrm{MPa}$ \\
 \hline
Mean Cell Poisson Ratio ($\nu_{i}$) & 0.5 \\
 \hline
 Death Rate ($k_d$) & $10^{-6} \mathrm{s^{-1}}$ \\
 \hline
Mean Receptor Concentration ($c^{rec}$) & 1.0 (Normalized) \\
\hline
Mean Ligand Concentration ($c^{lig}$) & 1.0 (Normalized)\\
\hline
Threshold Pressure ($p_c$) & $1.5\times10^{-7} \mathrm{MPa}$ \\
\hline
\end{tabular}
\label{table_1}
\caption{The parameters used in the simulation. For details, see \cite{malmi2018cell}.}

\end{table}

The net force (${\bf F}_i$) on the $i^{th}$ particle is the vectorial summation of elastic and adhesive forces that the neighboring particles exert on it (${\bf F}_i=\sum_{j=1}^{NN(i)}{\bf F}_{ij}$). Here, $j$ is summed over the number of nearest neighbors $NN(i)$. We performed over damped (low Reynolds number \cite{purcell1977life}) dynamics without thermal noise because the viscosity is assumed to be large. Therefore, the equation of motion for the $i^{th}$ particle is,
\begin{equation}
\dot{\textbf{r}}_i=\frac{{\bf F}_i}{\gamma_i},
\end{equation}
where $\gamma_i=6\pi\eta R_i$ is the friction term which models the environment as a thick gel and ${\bf r}_i$ is the position of the $i^{th}$ particle.

{\bf (b) Active Processes:} In the simulation the growth of individual particles is stochastic. They undergo division on reaching a critical radius ($R_m=5~\mu m$). The growth of the $i^{th}$ particle is dependent on the pressure $p_i$ due to neighboring particles. Therefore, the growth of individual particles are micro-environment dependent. We use Irving-Kirkwood definition \cite{irving1950statistical} to calculate pressure on the $i^{th}$ particle,   
\begin{equation}
    p_i=\frac{1}{V_{NN(i)}+ V_i}\sum_{j=1}^{NN(i)}{\bf F}_{ij}\cdot d{\bf r}_{ij}. 
\end{equation}
Here, $V_{NN(i)}= \sum_{j=1}^{NN(i)} \frac{4}{3}\pi R_j^3$ is the volume of nearest-neighbors of the $i^{th}$ particle, $V_i=\frac{4}{3}\pi R_i^3$ is the volume of the $i^{th}$ particle and $d{\bf r}_{ij}={\bf r}_i-{\bf r}_j $. 
If $p_i$ is smaller than a threshold value, $p_c$, the particles grow in size. However, if $p_i > p_c$, the $i^{th}$ particle becomes dormant with no size growth or division. Hence, the particles can switch between dormancy and growth mode depending on the ratio of $\frac{p_i (t)}{p_c}$. The volume of an individual particle grows stochastically in time and it divides into two daughter particles on reaching the critical radius $R_m$.  On division, two identical daughter particles are created with radii  $R_d=\frac{R_m}{2^{\frac{1}{3}}}$. Hence, a key time scale in the simulation is $\tau$ - the average time it takes for a particle to divide, set to be $\sim 15~$hours. 
Death of a particle take place in the simulations leading to a particle being randomly removed. The death rate is given by $k_d=10^{-6} s^{-1}$. Owing to $k_d << \frac{1}{\tau}$, we are simulating a rapidly growing system of particles.

{\bf Initial Conditions:} We initiated the simulations by placing 100 particles whose $x$, $y$, $z$ coordinates are chosen from a normal distribution with zero mean and standard deviation $30~\mu m$. In the present study, all the parameters have been fixed except the inter-particle adhesion strength $f^{ad}$ which is varied from $0$ to $3\times 10^{-4} \mu N/\mu m^2$. The simulated dense aggregate was evolved for $650,000s$ or $12\tau$. Relevant parameters are shown in Table~1. The time-dependent coordinates of particles were recorded in order to calculate the dynamical observables relevant to the present study.

\begin{center}
\begin{figure}[h]
\includegraphics[scale=0.12]{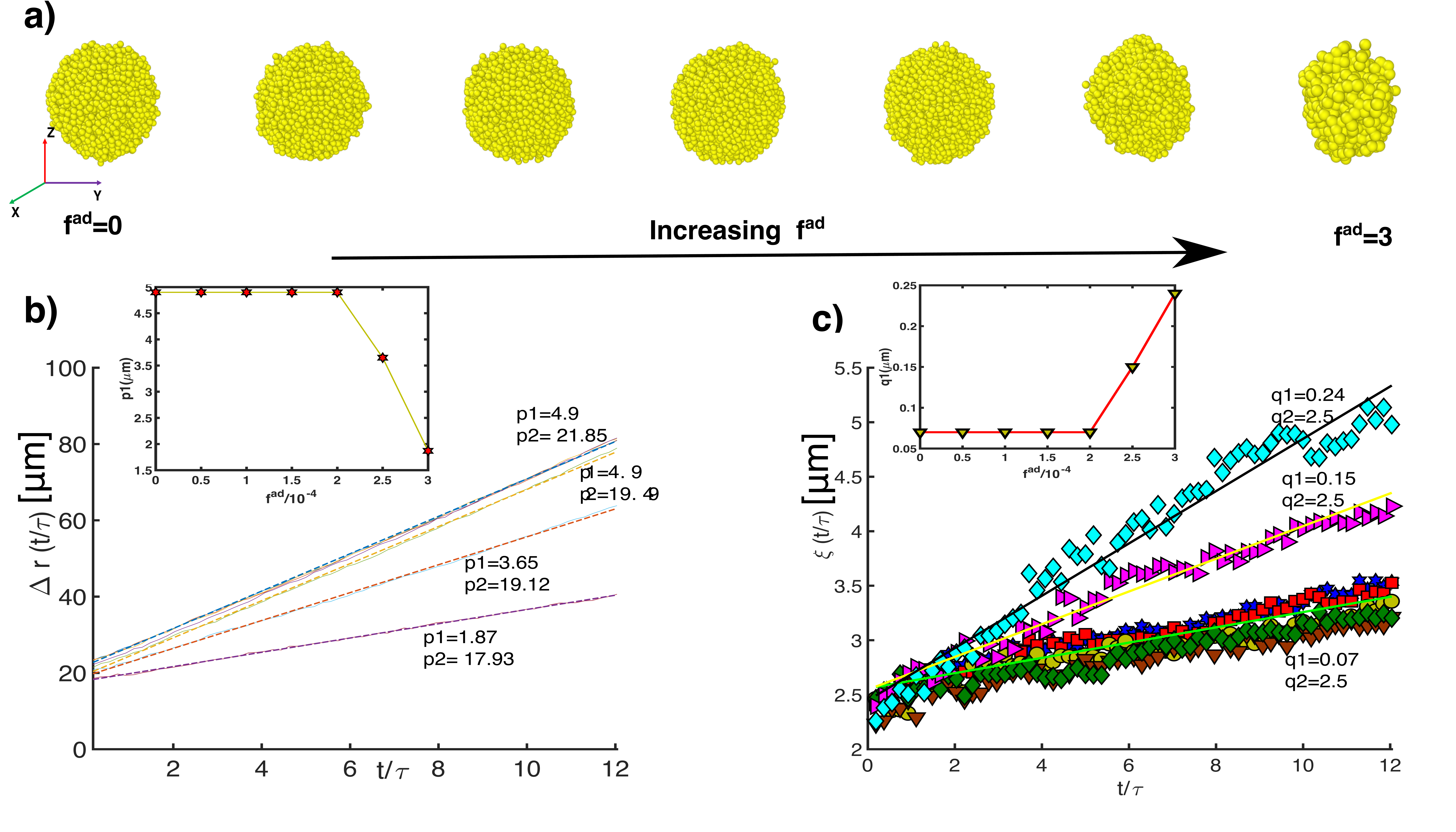}
\vspace{-.3 in}
\caption{{\bf Boundary front dynamics of the aggregate as a function of $f^{ad}$}. (a)Snapshots of aggregate at $12~\tau$ for different $f^{ad}$ values. From left to right, $f^{ad}/10^{-4}=0, 0.5, 1, 1.5, 2, 2.5$ and $3$ respectively. The graininess increases as $f^{ad}$ is increased.(b) Aggregate radius, $\Delta r$, as a function of time for different $f^{ad}$ values. The x-axis has been scaled by $\tau$. From top to bottom, $f^{ad}$ increases. $\Delta r(\frac{t}{\tau})$ has been linearly fitted to $p1(\frac{t}{\tau})+ p2$. $\Delta r$ corresponding to  $f^{ad}/10^{-4}=0, 0.5, 1, 1.5$, approximately have the same (p1, p2) values (i.e p1 and p2 are 4.9 and 21.85 respectively). For    $f^{ad}/10^{-4}=2, 2.5$ and $3$, (p1, p2) values are (4.9, 19.49), (3.65, 19.12) and (1.87, 17.93) respectively. Inset shows the dependence of $p1$ on $f^{ad}$.(c) Aggregate roughness, $\xi$, as a function of scaled time($\frac{t}{\tau}$) for different $f^{ad}$ values.The shapes on left correspond to the different $f^{ad}$ values. $\xi(t/\tau)$ was fitted to $q1(\frac{t}{\tau})+ q2$. $\xi (t/\tau)$ corresponding to  $f^{ad}/10^{-4}=0, 0.5, 1, 1.5$ and $2$ approximately have the same (q1, q2) values (i.e q1 and q2 are 0.07 and 2.5 respectively). For    $f^{ad}/10^{-4}= 2.5$ and $3$, (q1, q2) values are (0.15,2.5) and (0.24, 2.5 ) respectively. Inset shows the dependence of $q1$ on $f^{ad}$.}
\label{deltar_rough}
\end{figure}
\end{center}



\begin{figure}
\begin{center}
\hspace{-.2 in}
\includegraphics[scale=0.178]{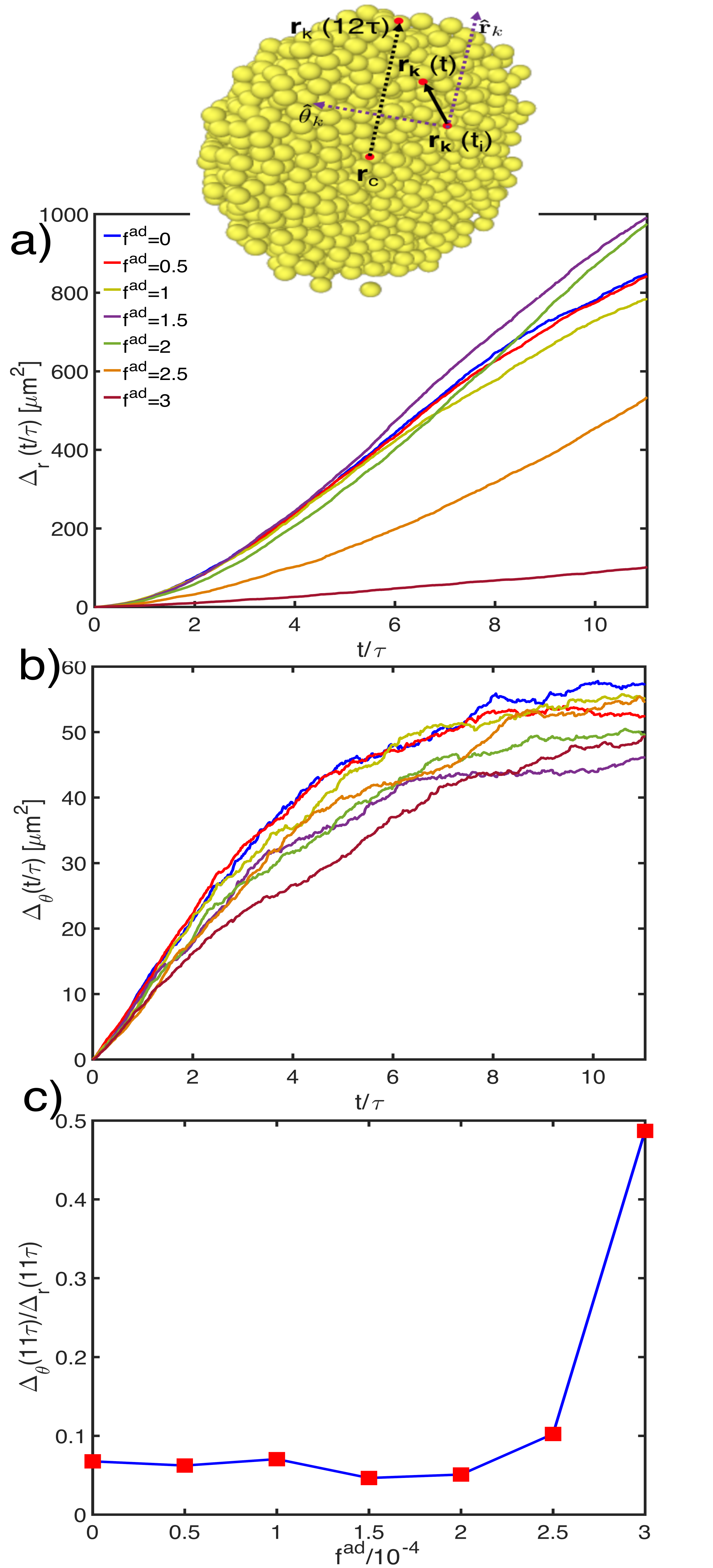} 
\vspace{-.2 in}
\hspace{.2 in}
\caption{{\bf Radial and transverse mean squared displacement for the boundary particles.} (a) Radial MSD, $\Delta_r(t)$, as a function of scaled-time ($\frac{t}{\tau}$) for different $f^{ad}$ values. The color-bars on the left correspond to the $f^{ad}$ values.  In the inset, we show a schematic how the radial ($\delta {\bf r}_k^{r} $) and transverse ($\delta {\bf r}_k^{\theta} $) displacement components for the $k^{th}$ cell was obtained. The $\hat{{\bf r}}_k$ is a unit vector parallel to ${\bf r}_k(12\tau)-{\bf r}_c$. The $\hat{\theta}_k$ is a unit vector perpendicular to $\hat{{\bf r}}_k$.(b) Transverse MSD ( $\Delta_{\theta}(t)$) as a function of scaled-time ($\frac{t}{\tau}$). The color of each curve (same as $\Delta_r(\frac{t}{\tau})$ ) corresponds to different $f^{ad}$ value. (c) Ratio of $\Delta_{\theta}(t^*)/\Delta_{r}(t^*)$ for $t^*=11\tau$. On increasing $f^{ad}$ beyond $2\times 10^{-4}$, $\Delta_{\theta}(t^*)/\Delta_{r}(t^*)$ increases sharply.  }
\label{rad_theta_msd}
\end{center}
\end{figure}

\section{Results}

\subsection{Inter-particle adhesion strength regulates boundary front dynamics}
Due to advancement in imaging modalities, boundary front dynamics of dense 3D aggregates such as in tumor cell collectives can be imaged~\cite{Valencia15, padmanaban2019cadherin}. The dynamical evolution of the boundary front is a collective observable, which as we show depends on the complex interplay between systematic interactions and active processes. In the present case, the boundary front dynamics is regulated by growth, division and death (active processes)  and inter-particle adhesion (systematic contribution). Figure \ref{deltar_rough}a shows snapshots of ensembles of simulated 3D dense active particle aggregates at fixed time $t=12\tau$ for increasing values of inter-particle adhesion strength, $f^{ad}$, from left to right. The aggregate snapshots at inter-particle adhesion strengths, $f^{ad}/10^{-4}=0, 0.5, 1, 1.5, 2, 2.5$ and $3$, show gradually increasing 3D interface roughness with smoother aggregate boundary at low values of $f^{ad}$. For $f^{ad}/10^{-4}=2.5$ and $3$, the aggregate boundary looks visually grainy. To probe the onset of roughness driven by increasing inter-particle adhesion strength, we first quantified the average boundary front dynamics using, 
\begin{equation}
    \Delta r (t)=\frac{1}{N_b(t)}\sum_{i=1}^{N_b(t)}|{\bf r}_i(t)- {\bf r}_c(t)|,
\end{equation}
where $\Delta r$ is the aggregate radius, $N_b(t)$ is the number of boundary particles at time $t$ and ${\bf r}_c(t)=\frac{1}{N(t)}\sum_{i=1}^{N(t)}{\bf r}_i(t)$ is the aggregate center for a total of $N$ particles. |..| denotes the absolute sign. As expected for a system with expanding number of particles, Figure \ref{deltar_rough}b shows the time dependent boundary growth ($\Delta r(\frac{t}{\tau})$) for $f^{ad}/10^{-4}=0, 0.5, 1, 1.5, 2, 2.5$ and $3$ (top to bottom). 
The boundary expansion is sensitive to the inter-particle adhesion strength with similar dynamics at  $f^{ad}/10^{-4}=0, 0.5, 1, 1.5, 2$ and marked differences starting to emerge for $f^{ad}/10^{-4}=2.5$ and  $3$. The boundary expansion is linear in time 
and is well fit by a 
linear function of the form $p1(\frac{t}{\tau})+ p2$ where $p1$ and $p2$ are fit coefficients. $\Delta r$ corresponding to  $f^{ad}/10^{-4}=0, 0.5, 1, 1.5$ approximately have the same (p1, p2) values (i.e p1 and p2 are 4.9 and 21.85 respectively). For higher inter-particle adhesion strengths $f^{ad}/10^{-4}=2, 2.5$ and $3$, (p1, p2) values are (4.9, 19.49), (3.65, 19.12) and (1.87, 17.93) respectively. As evident from the inset figure \ref{deltar_rough}b for $p1$ values as function of inter-particle adhesion strength ($f^{ad}$), the boundary expansion of the aggregate at $f^{ad}/10^{-4}=3$ is roughly three times suppressed compared to $f^{ad}=0$. These results show that on increasing $f^{ad}$, the spatial expansion of the aggregate is suppressed indicating that there is an interplay between systematic interactions and active processes. 

\subsection{Roughness of the dense aggregate is enhanced when radial and transverse motions are balanced}
The time dependent boundary expansion, $\Delta r$, discussed above gives an average picture of the surface dynamics but does not capture the fluctuations. The individual particle scale origin of fluctuations on the boundary front are of extreme relevance in the context of tumor progression and wound healing \cite{bru2003universal} in addition to the rich physics that underlie aggregate growth in view of the surface roughness~\cite{farrell2017mechanical}. Hence, we quantified the fluctuations of the aggregate boundary which we refer to as ``roughness'', $\xi$. The aggregate roughness, $\xi$, was calculated using, 
\begin{equation}
 \xi(t)=\frac{1}{N_b(t)}\sum_{i=1}^{N_b(t)}|d_i(t)-\Delta r(t)|,   
\end{equation}
where $d_i$ is the distance of the $i^{th}$ boundary particle from aggregate center (${\bf r}_c$). The 3D surface roughness increases linearly with time as the active particle aggregate spatially expands into the highly viscous medium at all values of inter-particle adhesion strength. 
We quantify the time evolution of the surface roughness ($\xi(\frac{t}{\tau})$) at multiple values of the inter-particle adhesion strength $f^{ad}/10^{-4}=0, 0.5, 1, 1.5, 2, 2.5$ and $3$ in Figure \ref{deltar_rough}c. 
By fitting $\xi(t/\tau)$ to a linear trendline  $q1(\frac{t}{\tau})+ q2$, we extract the roughness co-efficients $q1,q2$. 
The surface roughness co-efficients 
are approximately the same when $f^{ad}/10^{-4}=0, 0.5, 1, 1.5$ and $2$ as evident from (q1, q2) values (i.e q1 and q2 are 0.07 and 2.5 respectively). As the adhesion strength between active particles increases to $f^{ad}/10^{-4}= 2.5$ and $3$, a marked uptick in the surface roughness is observed with (q1, q2)  values equal to (0.15,2.5) and (0.24,2.5) respectively (see inset figure \ref{deltar_rough}c for the dependence of $q1$ on $f^{ad}$). We will show that the enhanced  aggregate roughness for inter-particle adhesion strength beyond $f^{ad}/10^{-4}=2$ is a nontrivial consequence of the dynamics of active particles at the surface of the 3D aggregate. Considering that the interface growth is driven by self generated particle flux, the relation between statistical properties of particle displacements and surface roughness is not well known. As we explain below, we relate the statistics of individual active particle displacements to the time evolution of the surface roughness.  

\subsection{Marked differences in radial and transverse individual particle dynamics suppresses 3D surface roughness}
To understand the enhanced surface roughness 
for $f^{ad}$ beyond $2 \times 10^{-4}$, we focus on the displacement of individual particles located at the aggregate boundary. We noted previously that the dramatic variations in the particle dynamics between the core and periphery of the aggregate~\cite{malmi2018cell, sinha2020spatially}, is a consequence of active processes being manifested predominantly on the surface of the aggregate. Consequently, we calculated the radial MSD ($\Delta_r(t)$) and transverse MSD ($\Delta_{\theta}(t)$) for particles whose distance from the aggregate center was greater than $\Delta r(t=11\tau)/2$ thus selecting for particles 
at the boundary. We define the transverse MSD ($\Delta_{\theta}(t)$) as,
\begin{equation}
\Delta_{\theta}(t-t_i, t_i)=\frac{1}{N}\sum_{k=1}^{k=N}[\delta {\bf r}^{\theta}_k(t-t_i, t_i)]^2,
\label{tmsd}
\end{equation}
where $\delta {\bf r}^{\theta}_k(t-t_i, t_i)$ is the transverse component of the displacement, $\delta {\bf r}_k(t-t_i, t_i)={\bf r}_k(t)- {\bf r}_k(t_i)$ (i.e perpendicular to the radial direction given by the vector $\hat{{\bf r}_k}=\frac{{\bf r}_k(12\tau)-{\bf r}_c}{||{\bf r}_k(12\tau)-{\bf r}_c||}$) (see inset of figure \ref{rad_theta_msd}a for a schematic). Here, $t_i= \tau$ and trajectories of particles were recorded between $\tau \leq t \leq 12\tau $ and in Eq. (\ref{tmsd}) $k$ is summed and averaged over all active particles at a distance from the aggregate center greater than $\Delta r(t=11\tau)/2$. Similarly, we define radial MSD ($\Delta_r(t)$) as, 

\begin{equation}
\Delta_{r}(t-t_i, t_i)=\frac{1}{N}\sum_{k=1}^{k=N}[\delta {\bf r}^r_k(t-t_i, t_i)]^2,
\end{equation}
where $\delta {\bf r}^{r}_k(t-t_i, t_i)$ is the radial component (parallel to the radial direction) of the displacement, $\delta {\bf r}_k(t-t_i, t_i)={\bf r}_k(t)- {\bf r}_k(t_i)$ (see inset of figure \ref{rad_theta_msd}a for a schematic). 
The ensemble averaged radial and transverse  mean-squared displacements are shown in Figures \ref{rad_theta_msd}a and \ref{rad_theta_msd}b respectively 
for $f^{ad}/10^{-4}=0, 0.5, 1, 1.5, 2, 2.5$ and $3$. Together, the two components of the mean-squared displacements reveal salient features of the particle dynamics on the surface of the 3D aggregate - (a) $\Delta_r$ decreases  considerably for $f^{ad}~>~2\times 10^{-4}$ (see main panel of \ref{rad_theta_msd}a), (b) 
Inter-particle adhesion strength has no prominent effect on the transverse mean-squared displacement.  
To quantify the relative contributions of the radial versus transverse particle movements in determining the surface roughness, 
we considered the ratio -  $\Delta_{\theta}(t^*)/\Delta_{r}(t^*)$ ($t^*=11\tau$) - between transverse and radial MSDs (Figure \ref{rad_theta_msd}c). 
The MSD ratio is roughly constant and small at $0\leq f^{ad}\leq 2\times 10^{-4}$, indicating that at low inter-particle adhesion strengths radial displacements outpaces transverse displacements. However, $\Delta_{\theta}(t^*)/\Delta_{r}(t^{*})$  dramatically increases beyond $f^{ad}=2\times 10^{-4}$  
with the ratio approximately an order of magnitude higher at $f^{ad}=3\times 10^{-4}$ compared to $f^{ad}=2\times 10^{-4}$. For $f^{ad}=2\times 10^{-4}$, $\Delta_{\theta}(t^*)/\Delta_{r}(t^{*})$ is $\approx 0.05$ whereas $\Delta_{\theta}(t^*)/\Delta_{r}(t^{*})$ is $\approx 0.5$ for $f^{ad}=3\times10^{-4}$. The larger ratio of the mean-squared displacements  
imply that the magnitude of transverse displacements become comparable to radial displacements in the regime of high inter-particle adhesion strength. This observation in the context of active particle aggregates is in agreement with polymeric systems the roughness of the surface increases when the transverse dynamics and radial dynamics are comparable~\cite{asai2018surface}. 

The emergence of comparable radial and transverse displacements at the surface of 3D aggregate 
can be anticipated through a physical argument. Aggregate boundary growth ($\Delta_{r}(t)$) is strongly governed by particle growth and division as it enables the radial expansion of the aggregate. As the aggregate expansion is  suppressed for $f^{ad}\geq 2\times 10^{-4}$ (see Figure \ref{deltar_rough}b) we surmise that the 
the relative contribution of systematic interactions as compared to active processes increase. 
Hence, transverse and radial displacements become comparable  $\Delta_{\theta}(t^*)\sim 0.5\Delta_{r}(t^{*})$, leading to enhanced 3D surface roughness. 

\begin{figure}
\includegraphics[scale=0.25]{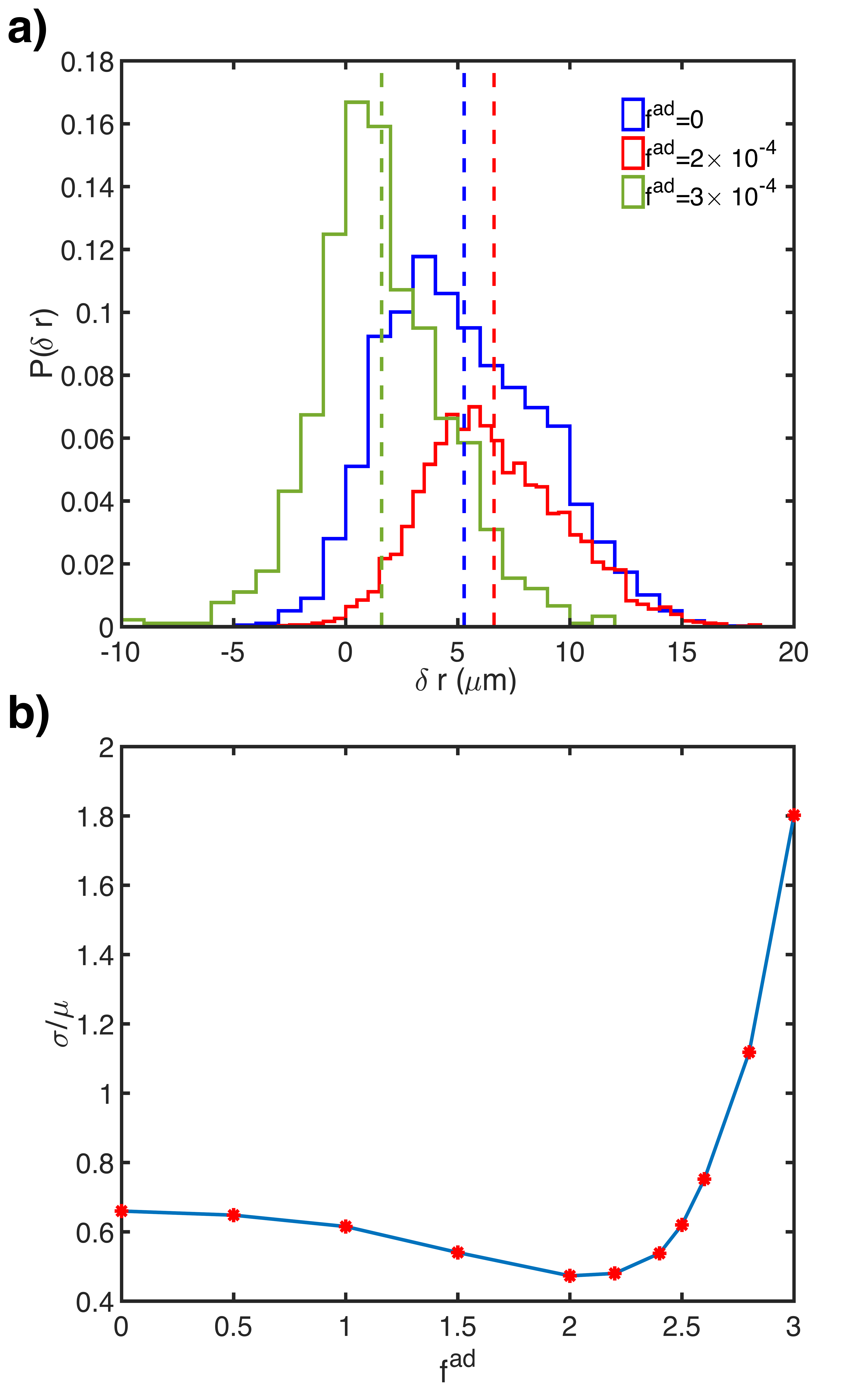} 
\vspace{-.2 in}
\caption{{\bf Coefficient of variation for radial displacements of particles} (a) Distribution ($P(\delta r)$) of radial component of displacement ($\delta r$). The dashed lines correspond to the respective mean of $P(\delta r)$. (b) Coefficient of variation ($\frac{\sigma}{\mu}$) for $P(\delta r)$ as function of $f^{ad}$. Here, $\sigma$ ($\mu$) is the standard deviation (mean) of $P(\delta r)$.  }
\label{rot_corr}
\end{figure}

\subsection{Heterogeneity in the orientation of radial displacements determine surface roughness}
Given the marked dependence of the radial MSD on inter-particle adhesion strength (Fig.~\ref{rad_theta_msd}a), we delved further into the statistical properties of active particle radial displacements and how surface roughness arises from  variations in the radial displacements of the active particles at the periphery of the 3D aggregate. 
How the spatio-temporal variations in the radial displacements of the particles at the 3D aggregate surface, what we refer to as the heterogeneity in radial displacements, relates to the surface roughness is not well known.  
To probe the aggregate surface fluctuations as motivated by our hypothesis, we calculated the 
radial component of displacements ($\delta r$) defined as, $\delta r_k= ({\bf r}_k(12\tau)-{\bf r}_k(9\tau))\cdot {\bf \hat{r}}_k$. Here, ${\bf \hat{r}}_k=\frac{{\bf r}_k(12\tau)-{\bf r}_c}{||{\bf r}_k(12\tau)-{\bf r}_c||}$ is a unit vector pointing from the center of the aggregate radially outward. 
The probability distribution, $P(\delta r)$, of radial displacement is calculated for all particles whose distance from the aggregate center was greater than $0.8 \Delta r(11\tau)$, selecting for boundary particles.  
We compare the probability distribution of the radial displacements in Figure \ref{rot_corr}a for $f^{ad}/10^{-4}=0, 2$ and $3$. 
Note that in the x-axis, negative values of the radial displacement $\delta r$ indicates particle movement towards the core while positive values indicate outward radial movement away from the center of the aggregate. 
The mean ($\mu$) of P($\delta r$) for $f^{ad}=3\times 10^{-4} \mu N/\mu m^2$ is considerably smaller than  $f^{ad}=2\times 10^{-4} \mu N/\mu m^2$ or $0$. 
The reduction in the mean radial displacement is due to a marked shift in the probability distribution towards negative values. Consequently, at higher inter-particle adhesion strengths, particles on the surface of the aggregate can either move outwards (away from the center; with positive radial movement $\delta r_k$) or inwards (towards the core; with negative $\delta r_k$) with comparable probability. To further quantify the heterogeneity in the radial displacements, we calculated the standard deviation ($\sigma$) or the spread of the distribution with changing inter-particle adhesion strength. 
Surprisingly, the standard deviation does not show a dependence on the inter-particle adhesion strength implying that the  variability in the active particle radial displacement by itself is not the main driver of surface roughness. However, the coefficient of variation ($\frac{\sigma}{\mu}$) of $P(\delta r)$ as a function of $f^{ad}$, 
is strongly enhanced at inter-particle adhesion strength $\ge 2.5\times 10^{-4}$ (see Fig.~\ref{rot_corr}b). Consistent with increasing roughness (Fig.~\ref{deltar_rough}c), the coefficient of variation in $P(\delta r)$ is higher with increasing inter-particle adhesion strength. Therefore, the roughness increases considerably on increasing $f^{ad}$ beyond $2\times 10^{-4} \mu N/\mu m^2$ due to the heterogeneity in the orientation of radial particle displacements - outward vs inward with respect to the core of the active particle aggregate. 
 
\section{Conclusion}
Surface dynamics of growing active matter aggregates is of crucial significance in understanding tumor invasion, wound healing as well as in furthering our understanding of interfacial stochastic growth. In this study, using a three dimensional active particle model, we delineated the role of particle-particle adhesion strength ($f^{ad}$) on aggregate boundary expansion and roughness. Our study reveals that the inter-particle adhesion strength controls the aggregate surface roughness by regulating the relative contributions of  radial and transverse of active particle movements on the periphery of expanding 3D aggregates. As inter-particle adhesion strength increases, the radial component of particle motion is significantly suppressed as observed from the reduced radial expansion of the 3D aggregate. The reduction in activity causes the systematic interactions to 
be prominent, leading to transverse and radial displacements being comparable. Subsequently, at high inter-particle adhesion strengths, particles at the 3D periphery can undergo radial displacements towards the core as well as radially outward thereby enhancing the surface roughness of the aggregate. 

The observation that roughness is controlled by the competition between radial and transverse components of particle motion has been reported in polymer grafted colloidal assemblies \cite{asai2018surface}. The emergence of a similar underlying principle  in active matter systems as compared to  conventional polymeric systems might pave way to the discovery of universal principles underlying the collective dynamics of active matter. The problem of active particle collectives expanding into a highly viscous medium provides a fascinating context for future studies into stochastic interface growth with particular relevance to biological systems such as tumor spheroids, organoids and bacterial aggregates. 
\section{Acknowledgement}
We would like to thank Prof. D. Thirumalai for discussions on the manuscript. We would also like to thank Xin Li and Himadri Samanta for their inputs during the course of this work. This work was supported by grants from the National Science Foundation (Grant Nos. PHY 17-08128 and PHY-1522550). AMK acknowledges support from start up funding at the College of Science and Mathematics, Augusta University. 
\section{Appendix}
Movies for the simulated particle aggregates. The total duration of the movie is $650,000~sec$ or $\approx 12 \tau$. The time interval between consecutive frames is $1000~sec$.\\
{\bf Movie 1:} Particle aggregates simulated for $f^{ad}=0~\mathrm{\mu N/\mu m^{2}}$. \href{https://drive.google.com/file/d/1vQTosnjMKXvxb4-2jO0P8py88SRxsqZT/view?usp=sharing}{(Link)}\\
{\bf Movie 2:} Particle aggregates simulated for $f^{ad}=2\times 10^{-4}~\mathrm{\mu N/\mu m^{2}}$. \href{https://drive.google.com/file/d/1BLOlwQDbUBJCAm84EjtknaaQKXm3z9I8/view?usp=sharing}{(Link)}\\
{\bf Movie 3:} Particle aggregates simulated for $f^{ad}=3\times 10^{-4}~\mathrm{\mu N/\mu m^{2}}$. \href{https://drive.google.com/file/d/1QUSpyiHtvSvkhAgM6gsfylXnVdXGHg5h/view?usp=sharing}{(Link)}. \\
\bibliographystyle{unsrt}
\vspace{-.3 in}
\bibliography{achemso-demo}

\end{document}